\documentclass[a4paper]{jpconf}
\usepackage{graphicx}

\newcommand{\LL}{\mathcal{L}}

\begin{document}
\title{Palatini approach to bouncing cosmologies and DSR-like effects}

\author{Gonzalo J. Olmo\footnote{Published under license in Journal of Physics: Conference Series by IOP Publishing Ltd.}}

\address{\footnotesize 	Departamento de F\'{i}sica Te\'{o}rica and IFIC, Centro Mixto Universidad de Valencia - CSIC.
  Facultad de F\'{i}sica, Universidad de Valencia, Burjassot-46100, Valencia, Spain.}

\ead{gonzalo.olmo@csic.es}

\begin{abstract}
It is shown that a quadratic gravitational Lagrangian in the Palatini formulation is able to capture different aspects of quantum gravity phenomenology in a single framework. In particular, in this theory field excitations propagating with different energy-densities perceive different background metrics, a fundamental characteristic of the DSR and Rainbow Gravity approaches. This theory, however, avoids the so-called {\it soccer ball problem}. Also, the resulting  isotropic and anisotropic cosmologies are free from the big bang singularity. This singularity avoidance occurs non-perturbatively and shares some similitudes with the effective dynamics of loop quantum cosmology. 
\end{abstract}

{\bf Introduction.} How could the Planck length, $l_P=\sqrt{\hbar G/c^3}\sim 10^{-35}$m, affect our understanding of space-time and quantum theory? To have a chance to answer this question, we should first understand how $l_P$ could be made compatible with the principles of relativity and general covariance. The tension between the coexistence of relativity and an invariant length scale is mainly due to our view that space-time assumes a Minkowskian continuous structure up to arbitrarily short scales. In this sense, to accommodate $l_P$ within our physical theories it may be useful to follow Riemann's advise and modify conventional geometric concepts as one applies them {\it both in the direction of the immeasurably large , and in the direction of the immeasurably small}. Einstein modified the space-time structure and dynamics at large scales, but kept the Minkowskian structure locally. Here it will be shown that a further revision of the classical space-time structure and dynamics at microscopic scales allows to introduce $l_P$ in an invariant and universal way compatible with general covariance. The resulting theory has interesting phenomenological aspects. On the one hand, we find that particles with different energy-momentum densities perceive different background geometries, a property closely related to doubly or deformed special relativity (DSR) \cite{DSR1,DSR2} and its gravitational extension, known as Rainbow Gravity (RG) \cite{Magueijo:2002xx}. On the other hand, the cosmological dynamics of the theory closely matches that of GR at all times except near the big bang singularity, which is replaced by a regular bounce that occurs near the Planck density. \\ 
 
{\bf Palatini approach to the idea of fundamental length.} Special relativity was built by requiring that the speed of light were an invariant and universal magnitude. To combine the speed of light and the Planck length in a way that preserves the invariant and universal nature of both quantities, we first note that $c^2$ has the dimensions of a squared velocity but that $c$ is not a 3-velocity. Analogously, we may see $l_P^2$ as an invariant with dimensions of length squared but $l_P$ not being a 3-length\footnote{Note that this idea is radically different from lattice-type approaches to the problem of a minimum length.}. Dimensional compatibility with a curvature suggests that $l_P^2\equiv 1/R_P$ could be introduced in the theory via the gravitational sector. The situation is not as simple as it may seem at first because an action like
\begin{equation}\label{eq:f(R)}
S[g_{\mu\nu},\psi]=\frac{\hbar}{16\pi l_P^2}\int d^4x \sqrt{-g}\left[R+l_P^2R^2\right]+S_m[g_{\mu\nu},\psi] \ ,
\end{equation}
where $S_m[g_{\mu\nu},\psi]$ represents the matter sector, contains $l_P$ but not in the invariant form that we wished. In fact, the field equations that follow from (\ref{eq:f(R)}) are equivalent to those of a Brans-Dicke theory with $\phi\equiv 1+2R/R_P$, a non-zero potential $V(\phi)=\frac{R_P}{4}(\phi-1)^2$, and $w=0$ \cite{Olmo2005}. As is well-known, in Brans-Dicke theory the observed Newton's constant is promoted to the status of field: $G_{eff}\sim G/\phi$. This allows the effective Newton's constant $G_{eff}$ to change in time and in space and, which makes $l_P^{eff}$ also vary in space and time. This is quite different from the assumed constancy and universality of the speed of light in special relativity and is closer to the philosophy of varying speed of light theories \cite{VSL}. The situation does not improve if we introduce higher curvature invariants in (\ref{eq:f(R)}) because they generate new degrees of freedom which turn Newton's constant into a dynamical field. \\
\indent A way out of this problem can be found by modifying the geometric structure of the theory. If we construct the theory {\it \`{a} la Palatini} \cite{review}, that is in terms of a connection not a priori constrained to be given by the Christoffel symbols, then the resulting  equations do not necessarily contain new dynamical degrees of freedom (as compared to GR), and the Planck length may remain space-time independent in much the same way as the speed of light. The field equations that follow from (\ref{eq:f(R)}) when metric and connection are varied independently are
\begin{equation}\label{eq:metric} 
f_R R_{\mu\nu}(\Gamma)-\frac{1}{2}f g_{\mu\nu}=\kappa^2 T_{\mu\nu} \ , \ \nabla_\alpha\left(\sqrt{-g}f_R g^{\beta\gamma}\right)=0 \ , 
\end{equation}
where $f=R+l_P^2R^2$, $f_R\equiv \partial_R f$, and $\kappa^2\equiv 8\pi G$. The connection equation can be easily solved after noticing that the trace of (\ref{eq:metric}), $R f_R-2f=\kappa^2T$, represents an algebraic relation between $R\equiv g^{\mu\nu}R_{\mu\nu}(\Gamma)$ and $T$, which implies that $R=R(T)$ and hence $f_R=f_R(T)$. For the particular Lagrangian (\ref{eq:f(R)}), we find that $R=-\kappa^2T$, like in GR. The connection turns out to be the Levi-Civita connection of an auxiliary metric $h_{\mu\nu}$ conformally related with the physical metric, $h_{\mu\nu}= f_R(T) g_{\mu\nu}$. We can now rewrite (\ref{eq:metric}) as follows
\begin{equation}\label{eq:Gmn-pal}
G_{\mu\nu}(h)=\frac{\kappa^2}{f_R(T)} T_{\mu\nu}+\Lambda(T)h_{\mu\nu} \ , \ \nabla_\alpha\left(\sqrt{-h}h^{\beta\gamma}\right)=0 \ ,
\end{equation}   
where $\Lambda(T)\equiv (f-R f_R)/(2f_R^2)=-(\kappa^2 T)^2 l_P^2$. Note that $\frac{\kappa^2}{f_R(T)}\equiv 8\pi G/(1-\kappa^2T/R_P)$  and the function $\Lambda(T)$ can be seen as a $T$-dependent Newton's constant and cosmological {\it constant}, respectively. This structure is very similar to that proposed in \cite{Magueijo:2002xx} for RG, $G_{\mu\nu}(E)=8\pi G(E)T_{\mu\nu}(E)+g_{\mu\nu}\Lambda(E)$, with the difference that the energy $E$ of the particle probing the geometry  is here replaced by $T$ which is a generally covariant function of its energy-density. From the structure of (\ref{eq:Gmn-pal}) and the relation $g_{\mu\nu}=(1/f_R) h_{\mu\nu}$, it follows that $g_{\mu\nu}$ is affected by the matter-energy in two different ways. The first contribution corresponds to the cumulative effects of matter, and the second contribution is due to the dependence on the local density distributions of energy and momentum. In fact, the structure of (\ref{eq:Gmn-pal}) is similar to that of GR, which implies that $h_{\mu\nu}$ is determined by integrating over all the sources (gravity as a cumulative effect).  Besides that, $g_{\mu\nu}$ is also affected by the local sources via the factor $f_R^{-1}(T)$. To illustrate this point, consider a region of the spacetime containing a total mass $M$ and filled with sources of low energy-density as compared to the Planck scale ($|\kappa^2 T/R_P|\ll 1$). For the quadratic model $f(R)=R+R^2/R_P$, in this region (\ref{eq:Gmn-pal}) boils down to $G_{\mu\nu}(h)=\kappa^2T_{\mu\nu}+O(\kappa^2T/R_P)$, and $h_{\mu\nu}\approx (1+O(\kappa^2T/R_P))g_{\mu\nu}$, which implies that the GR solution is a very good approximation . This confirms that $h_{\mu\nu}$ is determined by an integration over the sources, like in GR. Now, if this region is traversed by a particle of mass $m\ll M$ but with a non-negligible ratio $\kappa^2T/R_P$, then the contribution of this particle to $h_{\mu\nu}$ can be neglected, but its effect on $g_{\mu\nu}$ via de factor $f_R^{-1}=1-\kappa^2T/R_P$ on the region that supports the particle (its classical trajectory) is important. This phenomenon is analogous to that described in RG, where particles of different energies (energy-densities in our case) perceive different metrics. Our approach, however, naturally cures the so-called {\it soccer ball problem} for macroscopic bodies thanks to the dependence of the metric on the local energy-momentum densities rather than on the total energy-momentum of the particle. \\

{\bf Beyond $f(R)$ theories.} The energy-density properties of the Palatini theory (\ref{eq:f(R)}) is maintained with a much richer phenomenology when the $f(R)$ family is extended to Palatini  $f(R,Q)$ Lagrangians, where $Q\equiv R_{\mu\nu}R^{\mu\nu}$, and $R_{\mu\nu}$ is the symmetric Ricci tensor. In this case, the connection can also be expressed as the Levi-Civita connection of a metric $\tilde{h}_{\mu\nu}$ related with $g_{\mu\nu}$ by $\tilde{h}^{\mu\nu}=\frac{g^{\mu\alpha}{\Sigma_\alpha}^\nu}{\sqrt{\det \Sigma}}$ , where ${\Sigma_\alpha}^\nu=f_R\delta_\alpha^\nu+2f_Q {P_\alpha}^\nu$ and ${P_\mu}^\nu=R_{\mu\alpha}g^{\alpha\nu}$ 
are functions of ${T_\mu}^\nu=T_{\mu\alpha}g^{\alpha\nu}$ and, therefore, depend on the local densities of energy and momentum (see \cite{Olmo2011,OSAT09} for details). The
field equations for the metric can be written in compact form as
\begin{equation}
{R_\mu}^\nu(\tilde{h})=\frac{1}{\sqrt{\det\hat\Sigma}}(\frac{f}{2}{\delta_\mu}^\nu +\kappa^2 {T_\mu}^\nu) \ . \label{eq:f(R,Q)-metric}
\end{equation}
For a scalar field with kinetic energy $\chi\equiv g^{\mu\nu}\partial_\mu\phi \partial_\nu\phi$ and Lagrangian $\LL=\chi+2V(\phi)$,  
$\tilde{h}_{\mu\nu}$ and $g_{\mu\nu}$ turn out to be related by
\begin{equation}
g_{\mu\nu}=\frac{1}{\Omega} \tilde{h}_{\mu\nu}\ +\frac{\Lambda_2}{\Lambda_1+\chi\Lambda_2} \partial_\mu\phi \partial_\nu\phi  \label{eq:hdown}
\end{equation}
where $\Omega=\left[\Lambda_1(\Lambda_1+\chi\Lambda_2)\right]^{1/2}$, $\Lambda_1=\sqrt{2f_Q}\lambda+\frac{f_R}{2}$, $\Lambda_2={\sqrt{2f_Q}(-\lambda\pm\sqrt{\lambda^2+\kappa^2 \chi})}/{\chi}$, and $\lambda^2=f/2+f_R^2/8f_Q-\kappa^2\LL/2$. For the particular model 
$f(R,Q)=R-\frac{R^2}{2R_P}+\frac{Q}{R_P}$, the low energy-density limit $|\kappa^2\LL/R_P|\ll 1$ leads to
\begin{equation}\label{eq:LowLimit}
R_{\mu\nu}(\tilde{h})\approx \kappa^2\left(\partial_\mu\phi\partial_\nu\phi+\frac{V}{2}\tilde{h}_{\mu\nu}\right) 
+ \frac{\kappa ^2}{R_P}\left[({V -\tilde{\chi}})\partial_\mu\phi\partial_\nu\phi+\left(\frac{2\kappa^2 V^2+\kappa^2 \tilde{\chi}^2}{4}\right)\tilde{h}_{\mu\nu}\right] 
\end{equation}
which is in agreement with GR up to corrections of order $O(1/R_P)$. Note that to this order $\chi\approx \tilde{h}^{\mu\nu}\partial_\mu\phi \partial_\nu\phi\equiv\tilde{\chi}$. Similarly as in the $f(R)$ case, this indicates that $\tilde{h}_{\mu\nu}$ is mainly determined by integrating over the sources (cumulative effects of gravity), whereas $\Omega$ and the last term of (\ref{eq:hdown}) represent the local energy-density contributions to the metric.  The DSR limit of this theory can be obtained by neglecting the cumulative effects of gravity, which corresponds to the limit $\tilde{h}_{\mu\nu}\approx \eta_{\mu\nu}$. In this limit, for the  $f(R,Q)$ model of above the metric becomes
\begin{equation}
g_{\mu\nu}\approx \eta_{\mu\nu}+\frac{2\kappa^2}{R_P}\left(V\eta_{\mu\nu}+ \partial_\mu\phi \partial_\nu\phi\right) +O\left(\frac{1}{R_P^2}\right). \label{eq:hdown-Mink} 
\end{equation}
From (\ref{eq:hdown-Mink}) we see that the leading order corrections to the Minkowski metric are strongly suppressed by inverse powers of the Planck curvature, which indicates that a perturbative study of such contributions in field theories should be feasible at low energy densities (see \cite{Olmo2011} for more details). At high densities, however, non-perturbative effects arise. To illustrate this point, consider the scalar $Q$ in the $f(R,Q)$ model of above with a perfect fluid of density $\rho$ and pressure $P$
\begin{equation}\label{eq:Q-1/2}
Q=\frac{3R_P^2}{8}\left[1-\frac{2\kappa^2(\rho+P)}{R_P}+\frac{2\kappa^4(\rho-3P)^2}{3R_P^2} -\sqrt{1-\frac{4\kappa^2(\rho+P)}{R_P}}\right]\approx Q_{GR}+\frac{3 (P+\rho )^3}{2 R_P}+\ldots  \ .
\end{equation}
At low densities the GR solution is recovered but, as $\frac{4\kappa^2(\rho+P)}{R_P}$ approaches unity, important departures from GR must arise to avoid that the argument of the square root of (\ref{eq:Q-1/2}) becomes negative. This implies that $\rho$ and $P$ must be bounded if the dynamics is to be consistent. And this is so regardless of the symmetries of the theory. In the case of an isotropic FRW universe with constant equation of state $w=P/\rho$, the Hubble function in the spatially flat case is \cite{BO2010}
\begin{equation}\label{eq:Hubble-iso}
H^2=\frac{1}{6(\Lambda_1-\Lambda_2)}\frac{\left[f+\kappa^2(\rho+3P)\right]}{\left[1+\frac{3}{2}\Delta_1\right]^2} \ , 
\end{equation}
where $\Delta_1=-(1+w)\rho\partial_\rho \Omega$. At low densities, this expression exactly recovers the linear $\rho$-dependence of GR, 
but near the Planck scale, when $\kappa^2\rho/R_P\sim 0.1$, its behavior changes reaching a maximum and then vanishing at a higher density, which implies a cosmic bounce (see \cite{Olmo2011,BO2010} for illustrative plots). These isotropic bouncing solutions occur for all equations of state comprised within the interval $-1<w<11$ and persist in anisotropic (Bianchi I) spacetimes \cite{BO2010}. It should be noted that the effective dynamics of isotropic loop quantum cosmology \cite{lqc} for a massless scalar was exactly reproduced by an $f(R)$ Palatini theory in \cite{Olmo-Singh09}. The extension of that result to other matter sources and spacetimes with less symmetry could be naturally implemented in the context of $f(R,Q)$ theories.\\
\indent In this talk I have shown that without imposing any a priori phenomenological structure, Palatini models predict an energy-density dependence of the geometry that closely matches the structure conjectured by Magueijo and Smolin in \cite{Magueijo:2002xx,DSR2} but without the problems of that approach. At the same time, it avoids the big bang singularity in quite general situations and for reasonable sources of matter and energy such as dust and radiation. This confirms that Palatini theories represent a new and powerful framework to address different aspects of quantum gravity phenomenology. An important lesson that can be extracted from this analysis is that quantum gravitational phenomena associated with a minimum length scale may arise non-perturbatively, being the perturbative corrections strongly suppressed. \\
{\bf Acknowledgments.} Work supported by the Spanish grants FIS2008-06078-C03-02, FIS2008-06078-C03-03, and the Programme CPAN (CSD2007-00042).

\end{document}